\documentclass[11pt,twoside]{article}
\usepackage{asp2010}

\resetcounters

\aspvoltitle{Proceedings of ADASS XXI, Paris}
\aspcpryear{2012}
\aspvolume{???}
\def\urislash{\discretionary{/}{}{/}}
\def\uricolon{\discretionary{:}{}{:}}
{\catcode`\/\active
 \catcode`\:\active
 \catcode`\_\active
 \gdef\urli{\begingroup\ttfamily\fontsize{10}{12}\selectfont
    \catcode`\/\active \catcode`\_\active \catcode`\:\active
    \let/\urislash\let:\uricolon\def_{\char`\_}%
    \urlii}}
\def\urlii#1{#1\endgroup}

\begin{document}

\let\url\urli

\title{AstroDAbis: Annotations and Cross-Matches for Remote Catalogues}
\author{Norman Gray$^1$, Robert G Mann $^2$, Dave Morris$^3$, Mark Holliman$^2$ and Keith Noddle$^2$
\affil{$^1$SUPA School of Physics and Astronomy, University of Glasgow, Glasgow, G12 8QQ, UK}
\affil{$^2$Wide Field Astronomy Unit, Royal Observatory, University of Edinburgh,
  Edinburgh, EH9 3HJ, UK}
\affil{$^3$School of Physics,  University of Bristol,  Bristol, BS8 1TL, UK.}}

\begin{abstract}
Astronomers are good at sharing data, but poorer at sharing knowledge. 

Almost all astronomical data ends up in open archives, and access to
these is being simplified by the development of the global Virtual
Observatory (VO).  This is a great advance, but the fundamental problem
remains that these archives contain only basic observational data, whereas
all the astrophysical interpretation of that data -- which source is a
quasar, which a low-mass star, and which an image artefact -- is contained
in journal papers, with very little linkage back from the literature to
the original data archives.  It is therefore currently impossible for
an astronomer to pose a query like ``give me all sources in this data
archive that have been identified as quasars'' and this limits the effective
exploitation of these archives, as the user of an archive has no direct
means of taking advantage of the knowledge derived by its previous users.

The AstroDAbis service aims to address this, in a prototype service
enabling astronomers to record annotations and cross-identifications in the
AstroDAbis service, annotating objects in other catalogues.  We have
deployed two interfaces to the annotations, namely one astronomy-specific
one using the TAP protocol~\citep{std:tap}, and a second exploiting generic Linked Open
Data (LOD) and RDF techniques.
 \end{abstract}

\section{Introduction}
The AstroDAbis service provides a stand-off annotation service for
astronomical catalogue entries.  Catalogues appear in many forms, and
at scales ranging from tables in journal articles (later made
available electronically by the journals, in some cases) to large
software-engineering efforts on the part of specialised archives.  At
all scales, however, there are three key problems when working with
archives.

1. \emph{Catalogues contain information, but the knowledge derived
from analysis of them resides elsewhere, typically only in journal
articles}. This separation is intentional and well-motivated -- a
catalogue contains only values for readily measurable quantities, and
is, therefore, viewed as objective, while an astronomer's judgement
comes into play when those measurements are interpreted
astrophysically -- but the relationship between them is, typically,
asymmetric: through hyperlinks provided by ADS (\url{http://adsabs.harvard.edu/}) a journal
article can point to the online catalogue(s) used in its analysis, but
the data archives hosting the catalogue(s) do not implement analogous
pointers to the additional information about catalogue entries that
is present in the online literature.

2. \emph{Objects in the sky do not come with unique labels}. Combining
information from different catalogues requires \emph{cross-matching}.
This is expensive, and up to now the only way of speeding it up, by
either preserving the results of cross-matches, or providing
facilitating information on neighbour distances, has been to publish
(effectively) a new catalogue which, because it must be in an archive-specific format and
location, is hard to reuse widely.

3. \emph{Catalogues are static objects}.  Any such fresh catalogue,
derived from one or more existing catalogues but with the addition of
deduced or fresh information,  is logically independent from its progenitors.  Although there
is a chain of provenance, of course, the relationship of the new
information to the old is not available to the machine.

AstroDAbis addresses these three problems.
\begin{enumerate}
\item It provides a \emph{tagging} interface which allows users to
  associate annotations to catalogue objects.  This `folksonomy'
  tagging is inevitably imprecise, but (a)~this imprecision may be
  acceptable in some circumstances, and (b)~a very closely analogous
  mechanism will allow users to associate more semantically
  sophisticated annotations to objects, when consensus emerges on what
  such annotations should look like.
\item As a fundamental part of its design, the AstroDAbis service will
  implicitly create \emph{URI names} for every object in every
  catalogue it knows about.  This may seem profligate, or even
  impertinent, but (a)~the service will be able to declare equivalences to any URI
  names that a catalogue already supports, and (b)~as well as
  supporting cross-match tables, this creates the `raw materials' for
  other experiments deploying the Semantic Web within astronomy.
  Although it is not part of TAP at present, one could imagine an
  extension to TAP which documented a service's preferred pattern for
  URIs naming the objects it contains.
\item Stand-off tagging enables astronomer
users to annotate catalogues, and objects in catalogues, to which they
have no write access.  This creates the possibility of Web~2.0 or
Semantic Web infrastructures without requiring catalogues to make
the potentially disruptive changes to their systems which built-in
annotation would demand.  One implication of the Linked Data interface to
the service (see \url{http://linkeddata.org}) is that we provide RDF information about both the
catalogue objects, and the celestial objects they refer to, in a
flexible and open-ended way.
\end{enumerate}

The use of annotations to enrich existing data resources is not new to
science or to the Web world; indeed sites such as delicious.com or
Flickr are primarily concerned with such annotation in the form of
`tagging', and the associated notion of `folksonomy'.  Delicious-style tagging is one of the
inspirations of the AstroDAbis project but the more immediate one is
the Distributed Annotation System
(DAS, \url{http://www.biodas.org}), which is a widely-used
protocol for exchanging annotations on genomic and protein sequences.
This system inspired previous work involving one of us (Mann), on the
development of the
AstroDAS system~\citep{bose06}, which prototyped the recording
and publication of annotations of astronomical catalogues.
AstroDAS was a successful proof-of-concept, but
the immaturity of the VO protocol suite at that time meant that it
could not be implemented using open standards developed by the
IVOA, and this limited
its utility. Since then, the IVOA has released TAP, which
provides a standard means of accessing tabular astronomical datasets,
and important catalogues are becoming published to the VO through TAP
services.

At the time of writing the service is available as a prototype, but
as it matures (during 2012) we plan to bring it up to a supported service, hosted by
the Wide Field Astronomy Unit at Edinburgh.

\section{Using the service}

\subsection{TAP Factory and OGSA-DAI}

Although it is not a dependency, the AstroDAbis system was designed to
be naturally usable with a TAP Factory \citep{adassxxi:hume} based on
OGSA-DAI (a framework for distributed data and query management; see
\url{http://www.ogsadai.org.uk/}).  Using the TAP factory a service
provider can create a service, with a TAP interface, which allows a
user to make a TAP query which refers to multiple other TAP services.
The OGSA-DAI service then decomposes the query into a group of
single-service queries, and re-combines the result streams into a
single result set which it passes back to the end-user.  By this
means, and as illustrated in Table~\ref{t:annotation}, an astronomer
user can easily create a query which uses observational information
from one catalogue along with annotation, identification or neighbour
information from the AstroDAbis service.

\articlefigure[width=0.5\textwidth]{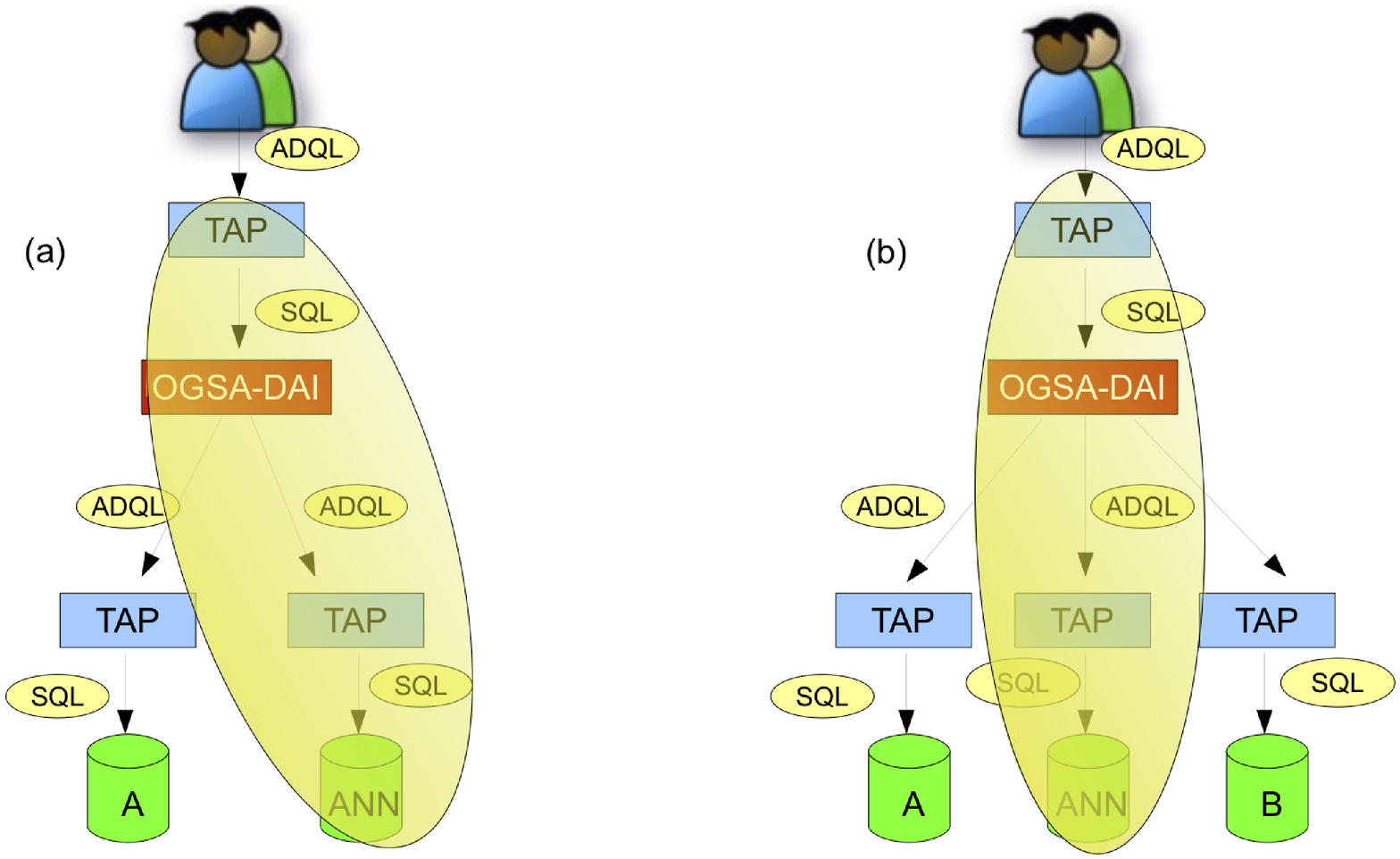}{f:ogsadai}{The OGSA-DAI
  architecture and AstroDAbis}

By providing a simple annotation service, the AstroDAbis mechanism
has the potential to support annotation of a very broad range of
astronomical objects, in a very broad range of repositories.

\subsection{Adding and retrieving annotations}

The service supports a web-based interface, which allows a user to
enter templated queries (which expand to ADQL queries~\citep{std:adql}), tagging the
objects which result.
Alternatively (and more suitably for batch-mode or bulk annotation),
users can upload annotations contained in a VOTable, as illustrated in Table~\ref{t:annotation}.
\begin{table}
\begin{verbatim}
SELECT TOP 100 masterObjID as pts_key,
  slaveObjID as objID, distanceMins as tagvalue
FROM twomass_pscXBestDR7PhotoObjAll
  <FIELD name='pts_key ID='masterObjID'
      ucd='meta.id;meta.main' datatype='long'>
    <DESCRIPTION>The unique ID in twomass_psc</DESCRIPTION>
  </FIELD>
  <FIELD name='objID' ID='slaveObjId'
      ucd='meta.id;meta.dataset' datatype='long'>
    <DESCRIPTION>The unique ID of the neighbour
      in BestDR7..PhotoObjAll (=objID)</DESCRIPTION>
  </FIELD>
  <FIELD name='tagvalue' ID='distanceMins' 
      ucd='pos.angDistance' datatype='float' unit='arcminutes'>
    <DESCRIPTION>Angular sep. between neighbours</DESCRIPTION>
  </FIELD>
\end{verbatim}
\caption{\label{t:annotation}An ADQL query which creates a VOTable,
  which can subsequently be uploaded to the AstroDAbis service to
  create a two-object annotation.}
\end{table}

Since the AstroDAbis service exposes a TAP interface to the world, its
annotation information is available through ADQL interfaces similar to the
one illustrated.

The TAP interface makes the AstroDAbis a first-class citizen in the
VO, so that its users' annotations can be combined with information
from other VO services to support high-level queries such as, for
example, ``find me the redshifts of all the objects which Fred Bloggs
identifies as quasars''.

As well as the TAP-based interface, AstroDAbis has a `Linked Data'
interface.  Although this provides utility by itself, it additionally
provides a mechanism for creating URI-based \emph{names}
for the objects in the catalogues it annotates.  These can act as a
springboard for future experiments with the Semantic Web in astronomy.

\subsection{Further information}

See \url{http://code.google.com/p/astrodabis/} for 
project source code and documentation.

\acknowledgements This work was funded by the Joint Information
Systems Committee (JISC), as part of its Content and Digitisation
Programme (\url{http://www.jisc.ac.uk/whatwedo/programmes/digitisation.aspx}).

\bibliography{P051}

\begin{thebibliography}{}
\expandafter\ifx\csname natexlab\endcsname\relax\def\natexlab#1{#1}\fi
\expandafter\ifx\csname url\endcsname\relax
  \def\url#1{\texttt{#1}}\fi
\expandafter\ifx\csname urlprefix\endcsname\relax\def\urlprefix{URL }\fi
\providecommand{\eprint}[2][]{\url{#2}}

\bibitem[{Bose et~al.(2006)Bose, Mann, \& Prina-Ricotti}]{bose06}
Bose, R., Mann, R.~G., \& Prina-Ricotti, D. 2006, in International Provenance
  and Annotation Workshop (I-PAW), edited by L.~Moreau, \& I.~Foster (Springer
  Verlag), vol. 4145 of LNCS, 154.
  \urlprefix\url{http://www.ipaw.info/ipaw06/proceedings/CameraReady_s7_4.pdf}

\bibitem[{Dowler et~al.(2010)Dowler, Rixon, \& Tody}]{std:tap}
Dowler, P., Rixon, G., \& Tody, D. 2010, Table access protocol ({TAP}, v1.0),
  {IVOA Recommendation}. \eprint{arXiv:1110.0497},
  \urlprefix\url{http://www.ivoa.net/Documents/TAP/}

\bibitem[{Hume et~al.(2011)Hume, Krause, Holliman, Mann, Noddle, \&
  Voutsinas}]{adassxxi:hume}
Hume, A.~C., Krause, A., Holliman, M., Mann, R.~G., Noddle, K., \& Voutsinas,
  S. 2011, in these proceedings

\bibitem[{Ortiz et~al.(2008)Ortiz, Lusted, Dowler, Szalay, Shirasaki,
  Nieto-Santisteban, Ohishi, O'Mullane, \& Osuna}]{std:adql}
Ortiz, I., Lusted, J., Dowler, P., Szalay, A., Shirasaki, Y.,
  Nieto-Santisteban, M.~A., Ohishi, M., O'Mullane, W., \& Osuna, P. 2008,
  {IVOA} astronomical data query language, version 2.0, {IVOA Recommendation}.
  \eprint{arXiv:1110.0503},
  \urlprefix\url{http://www.ivoa.net/Documents/latest/ADQL.html}

\end{thebibliography}
\end{document}